# Traveling-Wave Parametric Amplifier Based on Three-Wave Mixing in a Josephson Metamaterial

A. B. Zorin, M. Khabipov, J. Dietel, and R. Dolata
Physikalisch-Technische Bundesanstalt, Bundesallee 100, 38116 Braunschweig, Germany

*Abstract*—We have developed a recently proposed concept of a Josephson traveling-wave parametric amplifier with three-wave mixing. The amplifier consists of a microwave transmission line formed by an array of nonhysteretic one-junction SQUIDs. These SQUIDs are flux-biased in a way that the phase drops across the Josephson junctions are equal to 90°. Such a one-dimensional metamaterial possesses a large quadratic nonlinearity and zero cubic (Kerr-like) nonlinearity. This property allows phase matching and exponential power gain to take place over a wide frequency range. The proof-of-principle experiment performed at a temperature of $T$ = 4.2 K on Nb trilayer samples has demonstrated that our concept of a practical broadband Josephson parametric amplifier is valid and very promising for achieving quantum-limited performance.

## I. INTRODUCTION

Recently, parametric amplifiers based on the mixing of microwaves that travel in superconducting transmission lines with embedded discrete Josephson elements have been extensively studied [1–3]. This class of cryogenic devices, which have a large directional gain, a wide frequency bandwidth, and a large dynamic range, has already demonstrated a near-quantum-limited performance [1]. All of these Josephson traveling-wave parametric amplifiers (JTWPAs) operate on the four-wave mixing (4WM) principle, i.e., when the frequencies of the pump, of the signal and of the idler obey the relation $2f_p = f_s + f_i$. Such a regime is easily accessible due to the Kerr-like nonlinearity associated with a cubic term $\propto \varphi^3$ in the Taylor expansion of the Josephson current-phase relation $I(\varphi) = I_c \sin\varphi$, where $I_c$ is the critical current. However, the major obstacle to achieving a high flat gain in these 4WM JTWPAs is the need to precisely match the phases of the waves. Unfortunately, although cubic nonlinearity leads to desirable wave mixing, it also leads to undesirable self-phase and cross-phase modulations, causing dependence of the phase velocity on the wave power [4]. Therefore, sophisticated dispersion engineering is applied, with resonant elements added directly to the transmission line [5]. The disadvantages of this form of engineering are complexity of the JTWPA architecture and the unavoidable undulations of the gain with signal frequency [1,2].

It is, however, known that three-wave mixing (3WM) with frequencies that obey the traditional parametric relation $f_p = f_s + f_i$ is possible in both the optical medium and in the microwave transmission line [6] with noncentrosymmetric (e.g., quadratic) nonlinearity, and that this mixing is free of self-phase and cross-phase modulation effects. Moreover, in this case, the pump tone, $f_p \sim 2f_s$, is well separated from the amplified signal and can easily be filtered out. As was recently shown [7], it is possible to engineer such a nonlinear medium (Josephson metamaterial) that possesses quadratic nonlinearity and, therefore, to realize a JTWPA that is operated in advanced 3WM mode.

## II. NONCENTROSYMMETRIC NONLINEARITY

The easiest way to engineer quadratic nonlinearity in a Josephson circuit is to bias a Josephson junction (JJ) by a constant current, $|I_{dc}| < I_c$, thereby creating the constant phase drop $\varphi_{dc} = \arcsin(I_{dc}/I_c)$ (see, e.g., Ref. [8]). Then, the current phase relation $I(\varphi) = I_c \sin(\varphi_{dc} + \varphi)$ with respect to the small variation of phase $\varphi$ around the value $\varphi_{dc}$, $-\pi/2 < \varphi_{dc} < \pi/2$ takes the form

$$I(\varphi) \approx I_c \sin\varphi_{dc}(1-\varphi^2/2) + I_c \cos\varphi_{dc}(\varphi-\varphi^3/6). \quad (1)$$

Here, the term $\propto \varphi^2$ ensures the desired property, i.e., the current $I$ is no longer a centrosymmetric function of phase $\varphi$, i.e. $I(-\varphi) \neq -I(\varphi)$. A similar method for achieving a small quadratic nonlinearity by injecting a constant current has been recently applied to the superconductor kinetic-inductance based traveling-wave parametric amplifier [9,10]. These amplifiers originally exploit the small cubic nonlinearity of the kinetic inductance of superconducting wires. Although a broadband gain in the 3WM regime has been demonstrated [9], the admixture of a residual cubic (Kerr) nonlinearity led to multiwave mixing processes and prevented exponential growth of the signal power with growing waveguide length from being achieved [10].

A tunable, nonlinear Josephson element that allows the efficient control of both quadratic and cubic nonlinearities and, thereby enabling 3WM, has been proposed in Ref. [7]. This element comprises an inductively shunted JJ (i.e., a one-junction SQUID or an rf-SQUID) where the constant phase difference $\varphi_{dc}$ is set by an external magnetic flux $\Phi_e$ that is applied to the loop. The shunting inductance $L$, which is essentially a geometrical inductance of a superconducting wire (type A in Fig.1a) or a kinetic inductance of a short array of larger JJs (see type B in Fig.1b), should be smaller than the

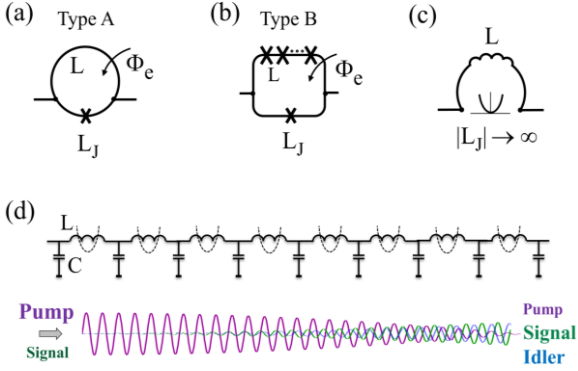

Fig. 1. Electric circuit diagram of the nonlinear element that has a one-junction SQUID configuration [7] based on (a) geometrical inductance (type A) and (b) kinetic inductance of a short serial array of $n$ large Josephson junctions with critical currents of $I_{cn} > nI_c$ (type B). (c) Equivalent circuit of the nonlinear element when external flux bias corresponds to the constant phase drop across the junction equal to $\pi/2$. Note that the Josephson junction adds to the resulting inductance only the nonlinearity (the Josephson current variation $\delta I \approx -0.5I_c\varphi^2$). (d) Electric diagram of the $LC$ transmission line including inductances with quadratic nonlinearity (upper panel) and schematic diagram showing the principle of the traveling-wave mixing resulting in a signal gain (bottom panel).

Josephson inductance of the prime junction $L_J(\varphi_{dc}) = \Phi_0/(2\pi I_c \cos\varphi_{dc})$ taken at $\varphi_{dc} = 0$, i.e. the screening SQUID parameter $\beta_L \equiv L/L_J(0) < 1$. This condition ensures the single-valued dependence of phase $\varphi_{dc}$ on flux $\Phi_e$. Thus, in contrast to the current-biased JJ, any bias value in the range $0 \leq \varphi_{dc} \leq 2\pi$ is accessible. In particular, when phase $\varphi_{dc} = \pm\pi/2$ the steady supercurrent circulating in the loop is equal to $I_c$ and the linear inductance of the junction $|L_J|$ is infinite. In this case, the junction does not contribute to the total linear inductance of the entire element (see Fig.1c), and all odd components of its nonlinearity (cubic, fifth, etc.) vanish as well. By contrast, the quadratic and higher-order even terms are large and the resulting current-phase relation takes the simple binomial form

$$I(\varphi)/I_c \approx \varphi/\beta_L - 0.5\varphi^2. \qquad (2)$$

The relative strength of the quadratic nonlinearity $|\chi| \equiv 0.5\beta_L$ approaches the ultimate value of 0.5 when parameter $\beta_L \to 1$.

### III. DESIGN

As proposed in Ref. [7], these discrete nonlinear elements form, together with corresponding ground capacitances, a one-dimensional array (Josephson metamaterial) that typically contains $N = 500 \div 1000$ elementary cells (see Fig.1d). The ground capacitors $C$ are designed in such a way that the wave impedance of the line $Z_0 = (L/C)^{1/2}$ is close to 50 Ω, whereas the cutoff frequency $f_0 = (LC)^{-1/2}/2\pi$ is reasonably low (e.g. ~ 50 GHz). Then, for a typical size of the elementary cell of $a \sim 30$ μm, the phase velocity is reduced, $v_0 = 2\pi f_0 a \sim 10^7$ m/s. The latter property allows a steep gain of the signal that propagates along the line (schematically shown in the bottom panel of Fig. 1d). For $f_s \sim 0.5 f_p$, the ultimate power gain is $G = \cosh^2(g_0 N)$, where the exponential gain factor $g_0$ is $|\chi|(\beta_L I_p/4I_c)(f_p/f_0)$ and $I_p$ is the ac pump current [7]. To make the chromatic dispersion of this line sufficiently low, the effective plasma frequency of the elementary cell, $f_c = (LC_J)^{-1/2}/2\pi$, where $C_J$ is the self-capacitance of the JJ, should be kept sufficiently high (i.e. ~ 100 GHz » $f_{s,i,p}$). In this case, the phase-matching condition $k_s + k_i = k_p$ (where $k_{s,i,p} \approx (f_{s,i,p}/f_0)(1 + 0.5 f^2_{s,i,p}/f_c^2)$ are the wave vectors of signal, idler, and pump, respectively) is approximately fulfilled over the whole length of the line. A detailed analysis of the problem of phase matching in the presence of small phase-modulation effects due to a residual Kerr nonlinearity has been carried out in Ref. [7], which also addresses the problems of pump depletion and signal-saturation power.

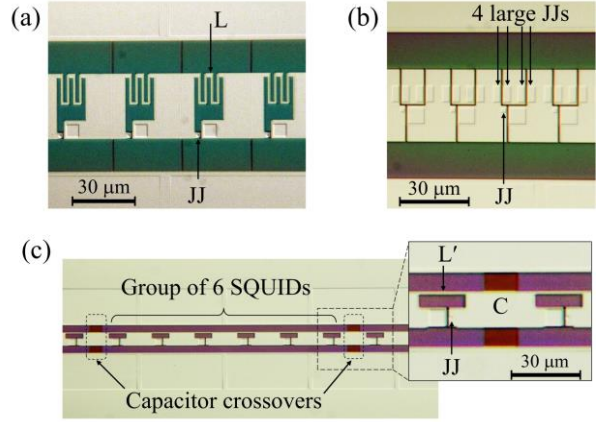

Fig. 2. Images of fragments of Nb JTWPAs with (a) elementary cells of type A, i.e. with a geometrical inductor realized as a meandered wire and (b) type B, i.e., with the kinetic inductance of a four-junction array. The latter design has a narrow loop; the necessary phase difference of $\pi/2$ is thus created by dc current $I_{dc}$ which is injected into the central conductor of the transmission line. Panel (c) shows a fragment of the JTWPA array that has elementary cells consisting of a group of 6 one-junction SQUIDs and one common plate capacitor $C$ formed by overlapping the metallic layers. The effective inductance of this cell $L$ is six times larger than the inductance $L'$ of an individual SQUID. For relatively large JJs with a low $L_J(0)$, this configuration ensures a sufficiently low cutoff frequency $f_0$ keeping $Z_0 \approx 50$ Ω. The dimensions of the small JJs are (a) 1 μm by 1 μm and (b) 2 μm by 2 μm, whereas the dimensions of large JJs are 3 μm by 7 μm, and (c) 1.4 μm by 1.4 μm. The largest quadratic JJs located next to the small ones are used with the purpose of layer inversion.

To check the operating principle of this JTWPA, we fabricated Nb circuits that comprised relatively large JJs with critical currents of $I_c \geq 10$ μA, which were designed to be operated at a liquid helium temperature of $T = 4.2$ K. The samples were fabricated using standard Nb/AlO$_x$/Nb trilayers with a nominal critical current density of 1 kA/cm$^2$, deposited on silicon wafers with a 300 nm SiO$_2$ layer on top. The details of our multilayer process, which allows JJs to be fabricated with linear dimensions down to 100 nm, are described in detail in Ref. [11]. Specifically, the process includes patterning using electron beam lithography and planarization by means of chemical mechanical polishing. The images showing different architectures of fabricated cells are presented in Fig.2.

## IV. EXPERIMENT

To measure the gain of our JTWPAs we used a simple setup for transmission-type measurements, which is schematically shown in Fig. 3a. To improve the signal-to-noise ratio we installed a cryogenic semiconductor preamplifier with a power gain of 40 dB, a sufficiency large bandwidth, and an effective noise temperature of about 5 K.

The sample measured consisted of an array of $N' = 1632$ one-junction SQUIDs divided into groups of $m = 6$ elements, each of which had the same ground capacitance (see Fig. 2c). Thus, this JTWPA essentially had $N = N'/m = 272$ elementary cells. This configuration was necessary due to the use of relatively large JJs with a critical current of $I_c = 12\,\mu\text{A}$ and, hence, a relatively small value of the Josephson inductance of $L_J(0) \approx 27\,\text{pH}$. The measured value of the geometrical inductance was 23 pH, yielding the value of the screening SQUID parameter $\beta_L \approx 0.85$. Combining these SQUIDs in groups of 6 elements yielded an effective inductance of each cell of $L \sim 138\,\text{pH}$. In combination with the effective ground capacitance $C \approx 55\,\text{fF}$ this inductance ensured a line impedance $Z_0$ close to $50\,\Omega$ and a cutoff frequency of $f_0 \approx 58\,\text{GHz}$. For an optimal pump power of $I_p \sim 0.5\,I_c$ we calculated the exponential gain factor $g_0 \approx 0.09$ and obtained a rough estimate of the power gain, $G \approx 15\,\text{dB}$.

The powers of the output signals measured in nondegenerate mode ($f_s = 6.4$ GHz $\neq f_i = 5.6$ GHz, and $f_p = 12$ GHz) are presented in Fig. 3b. The signal gain rises with rising pump power and approaches the level of $G_{max} = 11$ dB (it remains > 10 dB in the range from 4.8 GHz to 7.8 GHz). This value is somewhat smaller than the predicted gain of 15 dB. The discrepancy can be explained by inaccuracy in the estimation of the sample parameters and by the fact that our rough estimate was based on the analysis [7] which is valid in the limit of $I_p \ll I_c$.

We also measured the samples of type A, where each cell consisted of a single one-junction SQUID, and of type B, where the SQUIDs with JJ-based inductances were combined in groups of 6 elements. Both circuit types showed a robust performance with a reasonable gain, although their parameters were not close to optimal.

## V. CONCLUSION

In conclusion, the Josephson quadratic nonlinearity, which can be easily engineered with the help of superconductor technology, has a great potential. It is not limited by any of the several applications explored so far, including the JTWPA that is the subject of this paper and of [7], by the period-doubling bifurcation detector [8], or by the parametric amplifier based on the Josephson ring modulator [12]. We hold the view that superconducting circuits based on the proposed Josephson metamaterial, which possesses remarkable nonlinearity, will open up new opportunities for quantum optics using microwaves, including generation and amplification of nonclassical light. These circuits will therefore be particularly useful for quantum information processing.

### ACKNOWLEDGMENTS
We thank T. Weimann and R. Wendisch for their assistance in sample fabrication, H.-P. Duda for technical support of the experiment and P. Meeson and K. Porsch for the helpful discussions.

### REFERENCES

[1] C. Macklin et al., "A near-quantum-limited Josephson traveling-wave parametric amplifier," Science, vol. 350, pp. 307- 310, October 2015.

[2] T.C. White et al., "Traveling wave parametric amplifier with Josephson junctions using minimal resonator phase matching," Appl. Phys. Lett., vol. 106, pp. 242601-1-5, June 2015.

[3] M.T. Bell and A. Samolov, "Traveling-wave parametric amplifier based on a chain of coupled asymmetric SQUIDs," Phys. Rev. Applied, vol. 4, pp. 024014-1-7, August 2015.

[4] G.P. Agrawal, Nonlinear Fiber Optics, Academic Press, San Diego, 2007.

[5] K. O'Brien, C. Macklin, I. Siddiqi, and X. Zhang, "Resonant phase matching of Josephson junction traveling wave parametric amplifiers," Phys. Rev. Lett., vol. 113, pp. 157001-1-5, October 2014.

[6] A. L. Cullen, "Theory of the travelling-wave parametric amplifier," Proc. IEE B: Electron. Commun. Eng., vol. 107, pp. 101-107 , August 1959.

[7] A. B. Zorin, "Josephson traveling wave parametric amplifier with three-wave mixing," Phys. Rev. Applied, vol. 6, 034006-1-8, September 2016.

[8] A. B. Zorin and Y. Makhlin, "Period-doubling bifurcation readout for a Josephson qubit," Phys. Rev. B, vol. 83, pp. 224506-1-7, June 2011.

[9] M. R. Vissers et al., "Low-noise kinetic inductance traveling-wave amplifier using three-wave mixing," Appl. Phys. Lett., vol. 108, 012601-1-5, January 2016.

[10] R.P. Erickson and D.P. Pappas, "Theory of multiwave mixing within the superconducting kinetic-inductance traveling-wave amplifier", Phys.Rev. B, vol. 95, pp. 104506-1-27, March 2017.

[11] R. Dolata, H. Scherer, A.B. Zorin, and J. Niemeyer, "Single-charge devices with ultrasmall Nb/AlOx/Nb trilayer Josephson junctions," J. Appl. Phys., vol. 97, pp. 054501-1-8, February 2005.

[12] See the recent example of exploiting the Josephson nonlinear element of type B in: N.E. Frattini et al., "3-wave mixing Josephson dipole element," arXiv:1702.008.

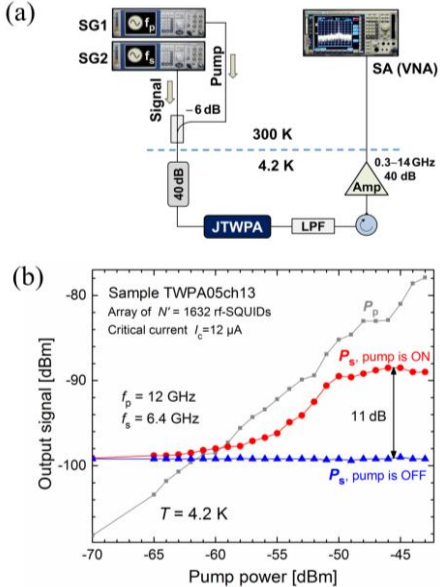

Fig. 3. (a) Schematic diagram of the transmission measurements performed at $T = 4.2$ K. The setup includes the sources of the pump (SG1) and of the signal (SG2), a spectrum analyzer (SA), a vector network analyzer (VNA), the cold HEMT preamplifier and the JTWPA sample. (b) Output signal and pump (idler not shown), measured in nondegenerate mode, at different levels of input pump power.